\documentclass[prf,amsmath,amssymb,floatfix,showpacs,longbibliography]{revtex4-1}

\usepackage{graphicx}
\usepackage{epstopdf, epsfig}
\usepackage{amsmath,amssymb}
\usepackage{braket}
\usepackage[english]{babel}

\newcommand{\beq}{\begin{equation}}
\newcommand{\eeq}{\end{equation}}
\newcommand{\vs}{\tilde{\bf v}}
\def\ku{ k_{_U} }
\def\lu{ \ell_{_U} }
\def\nus{ \nu_{_{3D}} }

\begin{document}

\title{3D instabilities and negative eddy viscosity in thin-layer flows}
\author{Alexandros \textsc{Alexakis}} 
\email[]{alexakis@lps.ens.fr}
\affiliation{Laboratoire de Physique Statistique, 
  \'Ecole Normale Sup\'erieure, PSL Research University; 
  Universit\'e Paris Diderot Sorbonne Paris-Cit\'e; 
  Sorbonne Universit\'es UPMC Univ Paris 06; CNRS; 
  24 rue Lhomond, 75005 Paris, France}
\date{\today}
\pacs{47.20.-k,
47.11.St,
47.11.Kb,
47.15.Fe,
}
\begin{abstract}
The stability of flows in layers of finite thickness $H$ is examined against small scale three dimensional (3D) perturbations and large scale two-dimensional (2D) perturbations.
The former provide an indication of a forward transfer of energy while the later indicate an inverse transfer and the possibility of an inverse cascade.
The analysis is performed using a Floquet-Bloch code that allows to examine the stability of modes with arbitrary large scale separation.
For thin layers the 3D perturbations become unstable when the layer thickness $H$ becomes larger than $H > c_1 (\nu \lu/U)^{1/2}= \lu Re^{-1/2}$, 
where $U$ is the rms velocity of the flown,  $\lu$ is the correlation length scale of the flow,  $\nu$ the viscosity and  $Re=\lu U/\nu$ is the Reynolds number.
At the same time large scale 2D perturbations also become unstable by an eddy viscosity mechanism when $Re>c_2$, where $c_1,c_2$ are order one non-dimensional numbers.
These relations define different regions in parameter space where 2D and 3D instabilities can (co-)exist and this allows to construct a stability diagram. 
Implications of these results for fully turbulent flows that display a change of direction of cascade as $H$ is varied are discussed.
\end{abstract}
\maketitle

\section{Introduction}        
\label{sec:intro}             

In many systems in nature there is a transfer of energy to both large and small scales \cite{Celani2010turbulence, Marino2013invers, pouquet2013geophysical, Seshasayanan2014edge,Deusebio2014dimentional, Seshasayanan2016critical, Sozza2015dimensional, Marino2015resolving, benavides2017critical, musacchio2017split}. This transfer at large Reynolds numbers leads to a split cascade of energy such that part of the energy cascades to the small scales and gets dissipated while at the same time part of the energy cascades to the large scales typically forming large scale coherent structures \cite{Byrne2011robust, seshasayanan2018condensates, alexakis2015rotatingTG, rubio2014upscale, Shats2010turbulence, Xia2011upscale, francois2013inverse,frishman2017turbulence}.
A typical example of such a behavior is observed in the Earth's atmosphere where, due to the combined effects of confinement, 
rotation and stratification, there is this split cascade leading to both small scale turbulence and fast mixing as well as large-scale structures 
like zonal flows and hurricanes. This split cascade has been quantified by in situ aircraft measurements in the hurricane boundary layer 
\cite{Byrne2013height}.
In the presence of a split cascade the amplitude of the inverse cascade depends on a control parameter, whose variation, makes the system transition
from a state exclusively cascading energy to the small scales to a state that the energy cascade is split or strictly inverse. 
The simplest example perhaps is turbulence in layers of finite thickness $H$ such that at scales $L$ much larger than the 
thickness, $L \gg H$,  the flow looks like two dimensional (2D) while at scales $\ell$ much smaller than the thickness, $H\gg \ell$,
the flow behaves like a three dimensional (3D) flow. Depending on the relative value of $H$ with respect to the forcing scale 
$\lu$ the system can behave like 3D cascading energy forward or like 2D cascading energy inversely. The amplitude of the 
inverse cascade appears to decrease as $H/\lu$ is increased until a critical value is met $H_c/\lu$ such that 
the inverse cascade becomes exactly zero. 

The mechanisms however under which such a coexistence of counter directed cascades exist are not yet understood.
In this work we try to unravel some of the mechanisms involved by examining the stability properties of simple laminar 
flows to both large and small scale perturbations. Although such an approach can not be used to study the full nonlinear
dynamics of a split cascade it provides a much simpler setup that can shed light on the mechanisms involved in the transfer of energy 
to small or larger scales. 
On the one hand it is known that 3D instabilities of laminar flows generate in general smaller scales transferring energy to them.
On the other hand 2D instabilities can couple the forced modes to large scale 2D modes transferring energy to larger scales. 
This non-local interaction of scales can be quantified with the use of an eddy-viscosity that can change sign depending 
on the flow parameters. A possible way to understand the mechanisms involved in the transition from a forward to an inverse cascade is to examine 
when 3D instabilities dominate and when there is a change of sign for the eddy viscosity of the flow. 

The notion of  eddy viscosity has been introduced in turbulence very early by Taylor \cite{taylor1915eddy}. First attempts for its calculation 
were made by Kraichnan \cite{kraichnan1976eddy} in an attempt to quantify the loss or gain of energy to the small scales. 
In this framework the evolution of a weak large scale flows $\bf v$ due to small scale turbulent fluctuations $\bf U$ can be described as
\begin{equation}
\partial_t {v^i} = - \sum_{j,k,m}\left( \nu \delta^{i,j}\delta^{l,m} + \nu_{eddy}^{i,j,l,m} \right) \nabla^l \nabla^m  v^j
\label{eddyvisc}
\end{equation}
where $\nu$ is the regular viscosity and $\delta^{i,j}$ stands for the Kronecker delta. 
The tensor $\nu_{eddy}^{i,j,l,m}$ is the eddy viscosity that models the effect of the small scale fluctuations $\bf U$ on $\bf v$.
Since the work of Kraichnan there have been many attempts to calculate an eddy viscosity for turbulent flows 
\cite{forster1977large, fournier1983remarks, yakhot1986renormalization, zhou1988renormalization, yakhot1989renormalization, zhou1989renormalized, germano1991dynamic}.
However the cascade processes of turbulent flows that excites a continuous spectrum of scales prevents
from having a closed expression for the eddy viscosity and even the notion of eddy viscosity for 
turbulent flows can be questioned. One case where the eddy viscosity tensor $\nu_{eddy}^{i,j,l,m}$  can be rigorously defined
is in the presence of a large scale separation between the forced small scale field $\bf U$ that evolves at scales $\lu$
and the flow velocity $\bf u$ that evolves at scales $L\gg \lu$. 
It can then be calculated in the low Reynolds number limit using homogenization theory \cite{dubrulle1991eddy,gama1994negative}. 
Here the Reynolds number is defined as 
$Re\equiv U \lu/\nu$ where                               %
$U$ is the rms of the small scale velocity field and     %
$\lu$ is the typical length-scale of the flow.           %
In the small Reynolds number limit and for isotropic flows it takes the form 
\begin{equation} \label{expeddyvisc}
 \nu_{eddy}^{i,j,l,m} = \frac{c_1}{\nu} U^2 \lu^2 \left[ \delta^{i,j}\delta^{l,m}  + \mathcal{O}\left(Re \right) \right] \quad \mathrm{for} \quad Re\ll 1,
\end{equation}
where $c_1$ is a non-dimensional number that depends on the detailed structure of the small scale field $\bf U$. 
For large $Re$ a viscosity independent value of $\nu_{eddy}$ is expected to be reached 
 \begin{equation} \label{expeddyvisc}
 \nu_{eddy}^{i,j,l,m} = c_2 U \lu \left[ \delta^{i,j}\delta^{l,m}  + \mathcal{O}\left(\frac{1}{Re}\right) \right] \quad \mathrm{for} \quad Re\gg 1.  
\end{equation}

If $-\nu_{eddy}^{i,j,l,m}$ has positive eigenvalues 
that are larger than the viscosity then the system can develop large scale instabilities \cite{cameron2016large}. 
In practice however it is not always feasible to calculate $\nu_{eddy}$ analytically, and calculations are limited to either very simple flows 
\cite{meshalkin1961investigation, nepomnyashchy1976stability,sivashinsky1985negative,sivashinsky1992negative}
or in the low $Re$ limit where the eddy-viscosity is sub-dominant to the regular viscosity.

In this work  a different path is followed by calculating the growth-rate of small and large scale instabilities using Floquet-Bloch theory
from which  the value of the eddy viscosity can be extracted. By doing so we can determine for the examined flows when 
the regions in the parameter space that the eddy viscosity is negative and large scale instabilities are present, and 
where the flow is dominated by three dimensional small scale instabilities instead.

\section{Formulation}
We begin by considering the Navier-Stokes equation for a unit density fluid in a triple periodic box of size ($2\pi L, \, 2\pi L,\, 2\pi H$)
with $L\gg H$ so that the layer has a small height compared to its length in the other directions.
\begin{equation}
\partial_t {\bf u } + {\bf u} \cdot \nabla {\bf u} = -\nabla P +\nu {\bf F} + \nu \Delta {\bf u},
\end{equation}
where $\bf u$ is the fluid velocity, $\nu$ is the viscosity and $\bf F$ is an external body force that maintains the flow. 
We assume the ${\bf F}$ is such that it supports a laminar time independent solution for the fluid $\bf U$. 
\[{\bf F }= -\nu \Delta {\bf U } + {\bf U} \cdot \nabla {\bf U} -\nabla P. \]
For simplicity a simple family of flows is going to be considered here, given by
\begin{equation}
{\bf U} =          U   \cos\left(\frac{r\pi}{2}\right)     \left[  \begin{array}{c} \sin( \ku y ) \\ \sin( \ku x ) \\ 0 \end{array} \right] + 
          \sqrt{2} U \,\sin\left(\frac{r\pi}{2}\right) \,  \left[  \begin{array}{c} \sin( \ku y ) \\ \sin( \ku x ) \\ 0 \end{array} \right] \sin( \ku z), 
          \label{eq:lam}
\end{equation}
where $\ku=1/\lu$ is the wavenumber of the flow and $U = {\|\bf U\|}$ is the $L_2$ norm of the flow.
Note that $U$ is independent of the parameter $r$. For the remainder of this investigation we will set $\ku=1$ and $U =1$.
With this normalization the Reynolds number of the flow is always given by 
\beq Re= \frac{U \lu}{\nu} = \frac{1}{\nu}. \eeq
For $r=0$, the flow is a two dimensional (2D) cell flow and is a solution of the Euler equations.
For $r\ne0$, $\bf U$ is three dimensional (3D) as it varies in all three dimensions.
For $r=1$ the flow reduces to a Taylor-Green flow \cite{TG1937mechanism}.
Note that if $r$ is not zero $\bf U$ is not a solution of the Euler equations and $\bf F$ has to have a more complex behavior than $\bf U$
to sustain it against viscosity and the non-linearities. Furthermore we note that for $r=0$, $H$ can take any value, while for $r\ne0$, 
$H$ is restricted to be an integer multiple of the flow period 
$H= n \lu$ with $n$ an integer.

The main objective of this work is to examine the stability of this flow and examine 
the evolution of an infinitesimal perturbation $\bf v$ to the laminar flow so that
${\bf u} = {\bf U} + {\bf v}.$
The linear evolution equation for $\bf v$ reads 
\begin{equation}
\partial_t {\bf v } + {\bf U} \cdot \nabla {\bf v} = -{\bf v} \cdot \nabla {\bf U} -\nabla P + \nu \nabla^2 {\bf v}.
\label{LNS}
\end{equation}
The linearity and homogeneity in time of the problem implies that asymptotically at large times $\bf v$ will have 
an exponential behavior with time and the goal is to determine the growth rate $\gamma$  of the perturbation in terms of the parameters of the system.
Because $\bf U$ is periodic in space and we are interested in the behavior of scales much larger than the scale of $\bf U$ 
it is convenient to use Floquet (or Bloch) theory. Floquet theory states that if $\bf U$ is periodic the perturbation field $\bf v$ can be decomposed as 
\begin{equation} 
{\bf v}({\bf x} ,t)=e^{i\bf q\cdot x} \vs({\bf x} ,t)+c.c.
\label{FLDC}
\end{equation} 
where ${\vs}({\bf x} ,t)$ is a complex vector field that has the same spatial periodicity as the velocity field $\bf U$, and $\bf q$ is an arbitrary wave number. The linear evolution equation for the field  ${\vs}({\bf x} ,t)$ then becomes
\begin{equation}
\partial_t \vs  + {\bf U} \cdot \nabla \vs + i {\bf U} \cdot {\bf q} \vs
                    = -\vs \cdot \nabla {\bf U} -\nabla \tilde{P} + \nu (\nabla+i{\bf q})^2 \vs.
                    \label{eq:floquet}
\end{equation}
The advantage of studying eq. (\ref{eq:floquet}) numerically as opposed to eq. (\ref{LNS}) is that one can consider
arbitrary large scale separations (determined by the vector parameter $\bf q$) with no additional computational cost.
Furthermore, the Floquet formulation gives a clear distinction between small scale and large scale instabilities.    
For $q=|{\bf q}|\ll \ku $, the volume average $\langle \vs \rangle$ over one spatial period $(2\pi \lu )^3$ gives the amplitude of $\vs$ at large scales $L\propto 2 \pi/{ q}$. 
Fields with $q=0$ and/or $\langle \vs \rangle=0$ correspond to purely small scale fields, 
and if these modes are unstable it amounts to a pure small scale instability {\it ie }: an instability that involves only wavenumbers that are integer multiples of the basic period $\ku$. 

Eq. (\ref{eq:floquet}) is solved numerically using a pseudospectral code. Details of the code can be found in \cite{cameron2016large,cameron2016fate}.
The resolutions examined varied from $32^3$ grid points for most of the results here although resolutions of $64^3$ grid points and $128^3$ grid points
were also used for large $Re$ and to test convergence. 

\section{Two dimensional flows $r=0$ }

We begin by examining the case for which $r=0$ so that the laminar flow is two-dimensional.
In this case the laminar flow can be written in terms of a stream function $\Psi= \cos(\ku x)-\cos(\ku y)$ as $U=[\partial_y \Psi, -\partial_x \Psi,0]$,
where the stream function is connected to the vertical vorticity $W$  as $W=-\nabla^2 \Psi= \ku^2 \Psi$.
The last equality holds because $\bf U$ contains modes of single wavenumber $\ku$.
The stability of this 2D flow will be examine in the next subsections, first against 2D perturbations
and second against 3D perturbations.  

\subsection{Two dimensional perturbations \label{2D2Ds}}

For velocity perturbations that are also 2D (ie $\partial_z \bf v =0$) the vertical component of the perturbation velocity decouples 
and follows a passive advection diffusion equation 
\begin{equation}
\partial_t {v_z } + {\bf U} \cdot \nabla {v_z} =   \nu \nabla^2 { v_z}.
\label{LNS}
\end{equation}
As a consequence its $L_2$ norm $\|{v_z } \|^2$ 
follows 
\[ \partial_t \|{v_z } \|^2 = -\nu \| \nabla v_z\|^2 \le -\nu \left(\frac{2\pi}{L}\right)^2 \| v_z\|^2\]  
and therefore the norm $\|{v_z } \|^2$ decays monotonically with time.
%
The remaining two components ${\bf v_{_{2D}}}=[v_x,v_y,0]$
can be written in terms of a stream function $\psi$ 
as $v_x=\partial_y \psi, v_y=-\partial_x \psi$. 
The evolution of these components can be written in terms of their vorticity $w=-\nabla^2 \psi$
as 
\begin{equation}
\partial_t {w } + {\bf U} \cdot \nabla {w} =-{\bf v_{_{2D}}} \cdot \nabla {W} -  \nu \nabla^2 {w}.
\label{LNS2D}
\end{equation}
Multiplying eq. (\ref{LNS2D}) by $w$ and space averaging we obtain after some manipulation (and using the laminar flow property $W=-\nabla^2 \Psi= \ku^2 \Psi$)
the enstrophy evolution equation 
\begin{equation}
\partial_t {\|w\|^2 } = \ku^2\langle w  {\bf U \cdot \nabla} \psi  \rangle -  \nu \| \nabla {w}\|^2.
\end{equation}
Multiplying eq. (\ref{LNS2D}) by $\psi$ and space averaging we obtain after similar manipulations
the 2D energy evolution equation
\begin{equation}
\partial_t {\|{\bf v}_{_{2D}}\|^2 } = - \langle \psi {\bf U} \cdot \nabla {w} \rangle  -  \nu \|  {w}\|^2.
\end{equation}
Multiplying the energy evolution equation by $\ku^2$ and subtracting from the enstrophy equation we obtain
\begin{equation} \label{bnd}
\partial_t ({\|w \|^2 - \ku^2 \|{\bf v}_{_{2D}}\|^2 }   ) 
= -  \nu   ( {\|\nabla w \|^2 - \ku^2\|w\|^2 }   ) 
\end{equation}
In general, the terms in parenthesis on the left and right hand side of the equation above \ref{bnd} can have either sign
and the equation above can not exclude the growth of $w$. 
If however $\ku=2\pi/L$ so that the domain size $L$ coincides with the velocity scale $2\pi \ku^{-1}$ 
({\it ie} no scales larger than the forcing scale are allowed)
both $(\|w \|^2 - \ku^2 \|{\bf v}_{_{2D}}\|^2)$ and $({\|\nabla w \|^2 - \ku^2\|w\|^2 })$ are
non-negative (by Poincar{\'e} inequality). The right hand side of eq. (\ref{bnd}) is then negative and the positive quantity $({\|w \|^2 - \ku^2 \|{\bf v}_{_{2D}}\|^2 }   )$ has to decrease
monotonically.  This implies that single scale 2D flows such that $\nabla^2 \Psi=-\ku^2 \Psi$ where $\ku$ is the largest scale of the domain are linearly stable to all 2D perturbations. 
In terms of the Floquet decomposition given in eq. (\ref{FLDC}) this conclusion translates to: the flow is stable for $q=0$ {\it ie} there are no small scale instabilities, and 
any unstable mode has to excite scales both smaller and larger than the forcing.  

Given this restriction we investigate numerically eq. (\ref{eq:floquet}) using a pseudo spectral code (described in \cite{cameron2016large,cameron2016fate})
and calculate the growth-rate $\gamma$ of the unstable modes 
as a function of $q$ and the viscosity $\nu$. The procedure to calculate this growth-rate is as follows.
We chose the wave-vector ${\bf q}=[q_x,0,0]$. In this case since we are interested in 2D perturbations $\bf q$ is restricted
in the $x$,$y$ plane. For simplicity, results are presented only for ${\bf q}=[q_x,0,0]$ that was found 
(but not proven) to be the most unstable from the general cases examined ${\bf q}=[q_x,q_y,0]$.
For this chosen wave-vector ${\bf q}=[q_x,0,0]$ the complex velocity field $\tilde{\bf v }$ is initialized using random initial conditions. 
Then the complex field $\tilde{\bf v }$ is evolved based on eq. \ref{eq:floquet} until a clear exponential growth 
of the energy $E=\langle |\tilde{\bf v }|^2 \rangle \propto e^{2\gamma t}$ is observed. 
The growth-rate $\gamma$ is then the calculated by fitting. This procedure is then repeated
for different values of $q$ and different values of viscosity. This allows to calculate the growth-rate
of the most unstable mode as a function of $q$ and $\nu$. It is noted that this procedure reveals only the fastest growing unstable mode 
and not all unstable modes in the system. It is worth keeping this in mind in the following sections where general 3D perturbations will be considered. 

\begin{figure}[!ht]
  \centering
  \includegraphics[width=0.45\textwidth]{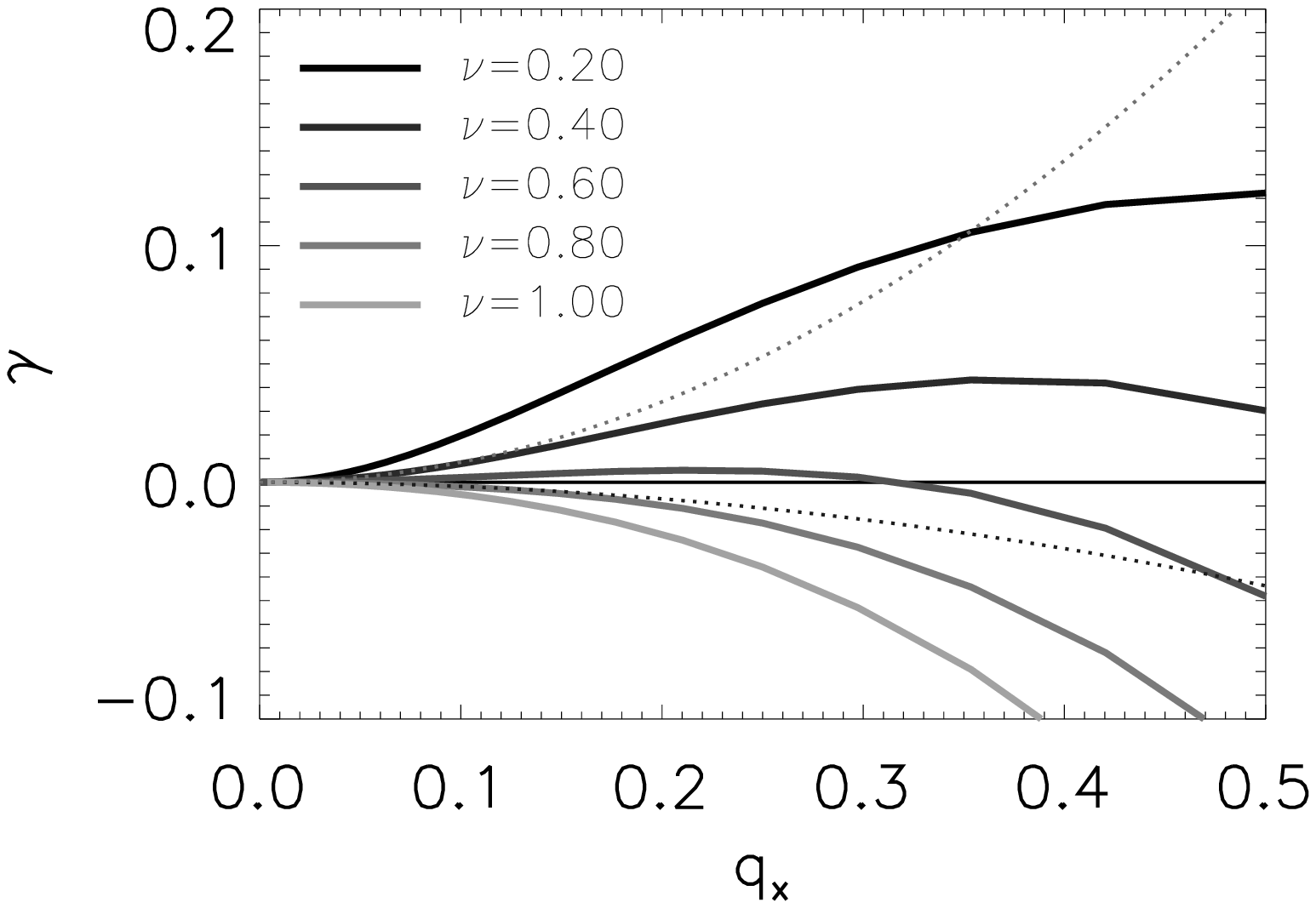}
  \includegraphics[width=0.45\textwidth]{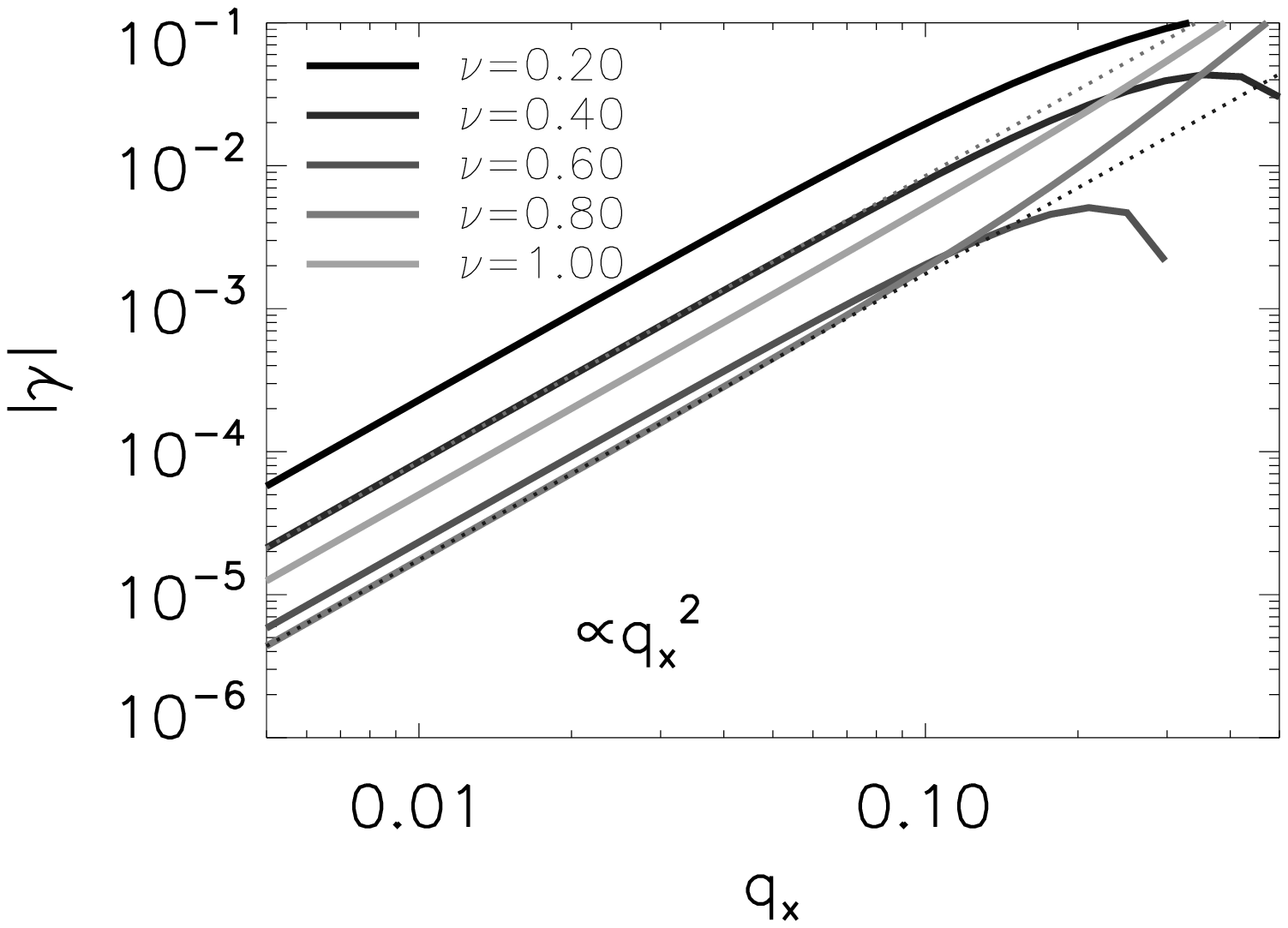}
  \caption{ The growth rate $\gamma$ as a function of the wavenumber $q$ for five different values of the the viscosity $\nu$.
            The dotted lines show fits to a quadratic power-law $\gamma= a q^2$. Left panel: linear scale. 
            Right panel: The absolute value of $\gamma$ in log log scale.    }
  \label{fig:2Dgrowth}
\end{figure}
The resulting growth rate $\gamma$ as a function of $q$ for different values of the viscosity is shown in the left panel of figure \ref{fig:2Dgrowth} in linear scale.
The right panel shows the absolute value of the same data in logarithmic scale. 
For small values of $\nu$ the growth rate peaks at $q_x =1/2$.
As discussed  in the beginning of this section for $q=0$ there can be no instabilities and the growth rate becomes zero as $q\to 0$.
The growth rate $\gamma$ is approaching zero as $q\to 0$ following a quadratic power-law. This is demonstrated 
more clearly in logarithmic scale shown in the right panel of figure \ref{fig:2Dgrowth} where it is fitted as 
\begin{equation} 
\gamma = a\, q^2
\label{q2}
\end{equation}
shown by the dotted lines.
The proportionality coefficient $a$ is negative for small values of the viscosity, while above a critical value it becomes positive.  
This particular scaling of the growth-rate $\gamma$ with the wavenumber $q$ implies that the instability is of the form of an eddy-viscosity. 
The eddy viscosity is a tensor and the proportionality coefficient $a$ in eq. (\ref{q2}) indicates the maximum eigenvalue
of $- \left( \nu \delta_{i,j} + \nu_{eddy}^{i,j,x,x} \right)$.
For simplicity $a+\nu$ will be denoted as $-\nu_{eddy}$ and referred as eddy viscosity although it must not be forgotten that
the eddy viscosity is truly a tensor. The relation of $\nu_{eddy}$ with $a$ is \beq a=-(\nu + \nu_{eddy}). \eeq 
With this notation a large scale instability is implied if $\nu_{eddy} < -\nu$.

\begin{figure}[!ht]
  \centering
  \includegraphics[width=0.5\textwidth]{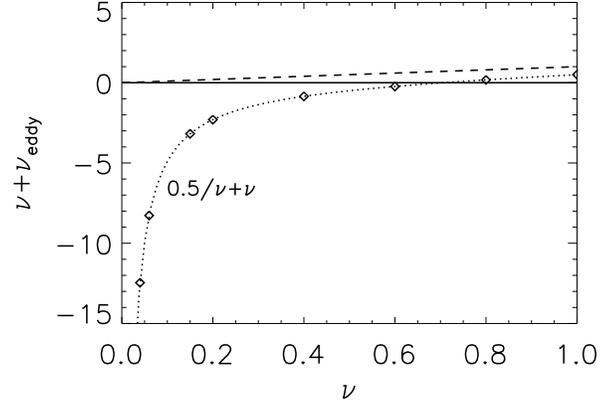}
  \caption{ The value of $\nu_{eddy}+\nu$ as a function of $\nu$.
  When this quantity takes negative values there is a large scale instability.
  The dashed line indicates where $\nu_{eddy}=0$. The doted line shows the fitting  
  to the asymptotic result $\nu_{eddy} = 0.5 \nu$. 
  Large scales become unstable when $\nu < \nu^*\simeq 0.7$. }
  \label{fig:2DeddyVisc}
\end{figure}
The eddy viscosity depends on the small-scale flow examined and the value of viscosity.
Figure \ref{fig:2DeddyVisc} shows the value of $\nu + \nu_{eddy}$ as a function of $\nu$ for the examined flow.
The dotted lines indicates the prediction (\ref{expeddyvisc}) for $c_1=0.5$. 
Negative values of $\nu + \nu_{eddy}$ imply that the flow amplifies large scale flows. 
For large values of $\nu$ the value of the eddy-viscosity is such that $\nu + \nu_{eddy}$ is positive
although still smaller than $\nu$. So although $\nu_{eddy}$ is negative and the flow reduces the decay rate caused by regular viscosity
it is not sufficiently strong to drive any large scale scale instabilities. Bellow some critical value of the viscosity $\nu = \nu^*$.
$\nu + \nu_{eddy}$ becomes negative. 
For values of $\nu$ smaller than $\nu^*$ the large scales of the flow are unstable. The critical value of the viscosity $\nu^*$ can be estimated to be 
\beq \nu^*\simeq 0.7 \,\, U \lu \label{nucr1} \eeq  
where we have kept $U=1$ and $\ku=1$ in the expression to recover the dimensions.

\begin{figure}[!ht]
  \centering
  \includegraphics[width=0.45\textwidth]{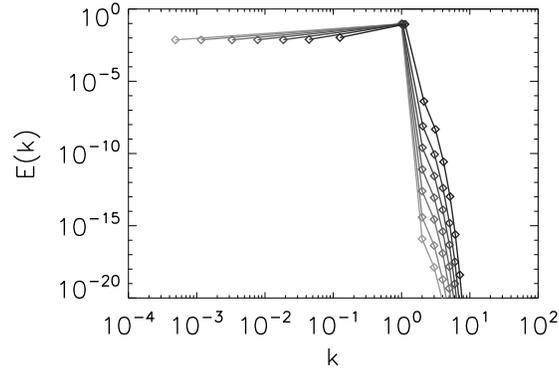}
  \caption{ Energy spectrum of the unstable modes $e^{i\bf q\cdot x} \vs({\bf x} ,t) $ for different values of $q$ and $\nu=0.2$, and $\ku=1$ 
            is the wavenumber of the laminar flow. The value of $q$ is indicated by the furthest point to the left. }
  \label{fig:spec2D}
\end{figure}
\begin{figure}[!ht]
  \centering
  \includegraphics[width=0.45\textwidth]{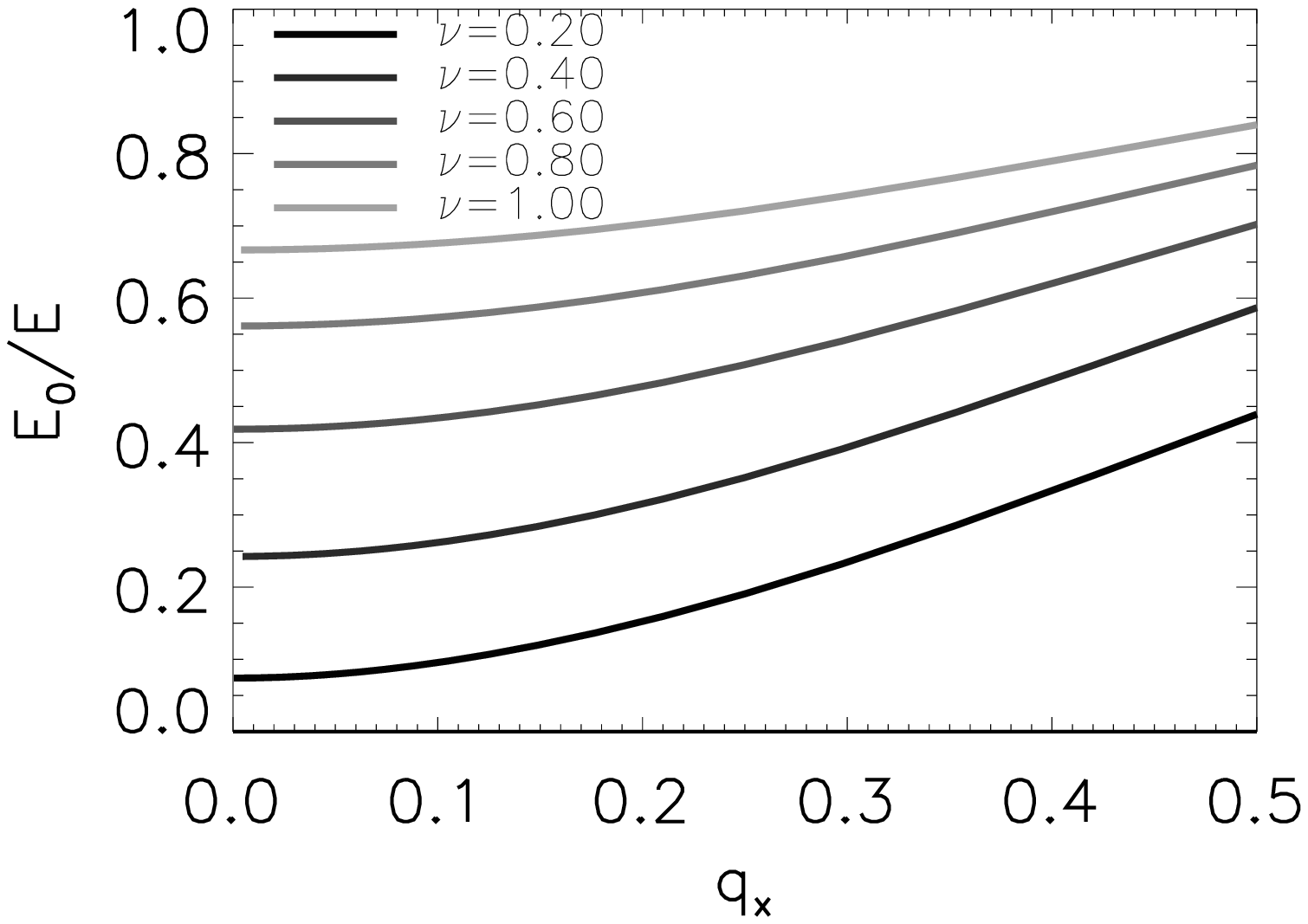}
  \includegraphics[width=0.45\textwidth]{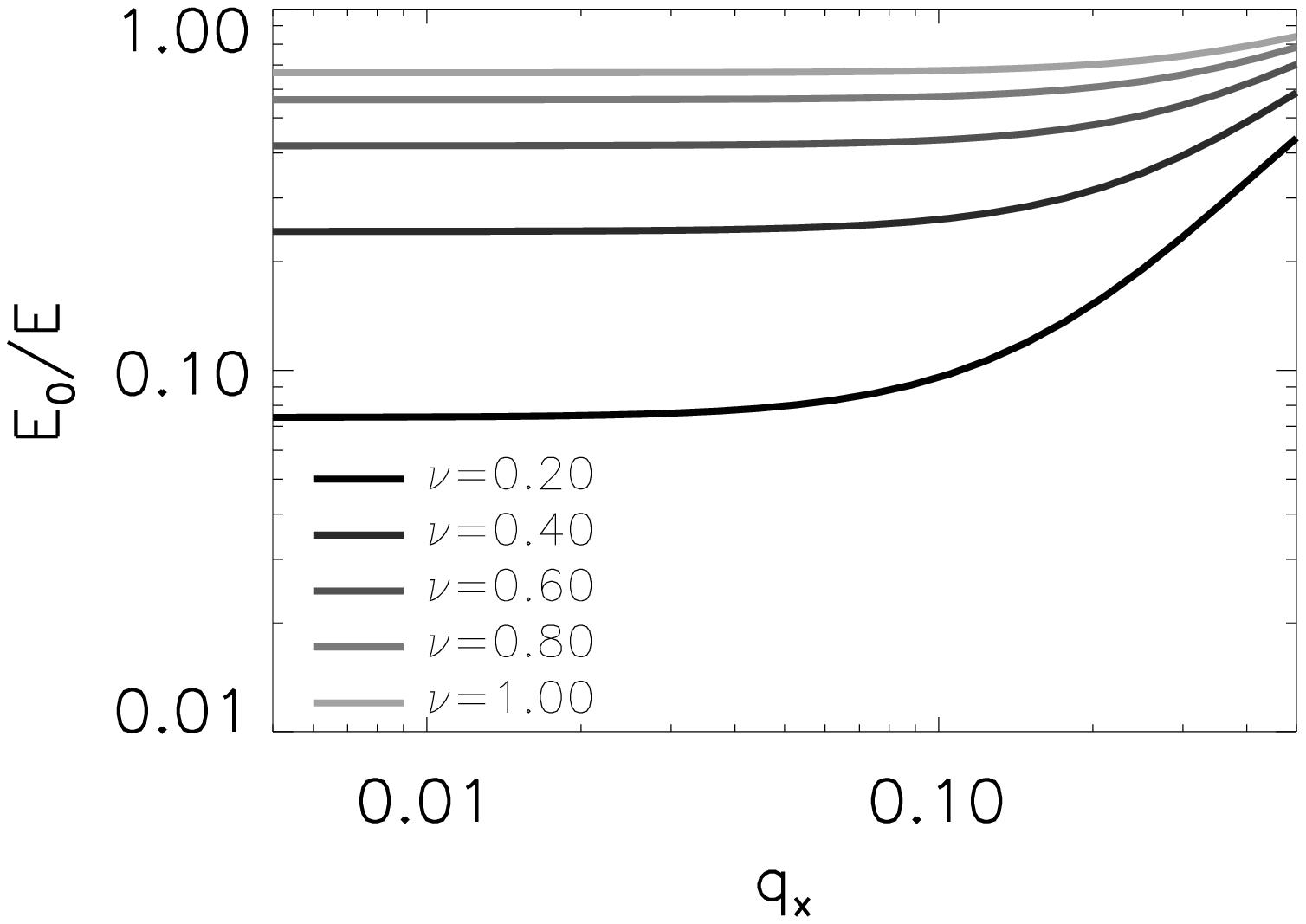}
  \caption{The ratio of the energy in the large scales mode  $E_0=\frac{1}{2} |\langle  \vs    \rangle|^2$ 
                                         to the total energy $E  =\frac{1}{2}  \langle |\vs|^2 \rangle   $
                                        as a function of $q$ for different values of  $\nu$. Left: on a linear scale, right in log-log scale.}
  \label{fig:Eratio2D}
\end{figure}

Before concluding this section it is worth examining the spectral shape of the unstable modes. 
Figure \ref{fig:spec2D} shows the energy spectra of the unstable modes as a function of the wavenumber
for different values of $q$. The spectrum is composed by the large scale component located at the wave number $q$ plus 
the part localized at small scales $k\ge \ku=1$. The amplitude of the energy at large scales is independent on
the value of $q$. This is seen in the figure \ref{fig:Eratio2D} where the ratio of the energy contained in the largest scale 
$E_0=\frac{1}{2}|\langle \vs \rangle|^2$ to the total energy $E=\frac{1}{2}\langle |\vs|^2\rangle$ is shown.
The amplitude of this ratio is initially decreased as $q$ is decreased but asymptotes to a $q$ independent value
for sufficiently small values of $q$. This asymptotic value decreases as $\nu$ decreases.

\subsection{Three dimensional perturbations \label{2D3Ds} }  

In the previous section we examined the stability of the 2D flow against large scale 2D perturbations. 
In this section we investigate the stability of the flow against 3D perturbations. This question is important because
it determines when 3D variations enter the system that can transfer energy to the small scales. 

For $r=0$ the flow $\bf U$ has no dependence on the $z$ direction. We can therefore consider an a layer with arbitrary thickness compared to the flow length-scale $\ku^{-1}$. 
The homogeneity in the $z$-direction also implies that each $q_z$ mode can be considered independently for the linear problem
with  the perturbation field $\vs$ being a function of $x,y,t$ only and not of $z$.
This section is restricted to 3D perturbations with wavenubers ${\bf q}=[0,0,q_z]$.
For this choice the field $\vs$ satisfies the incompressibility condition \beq \partial_x  \tilde{v}_x + \partial_y \tilde{v}_y + i q_z \tilde{v}_z =0.\eeq 
Note that unlike the previous case of 2D perturbations (with $q_z=0$)  the incompressibility condition implies that the $z$-component does not act independently.

\begin{figure}[!ht]
  \centering
  \includegraphics[width=0.45\textwidth]{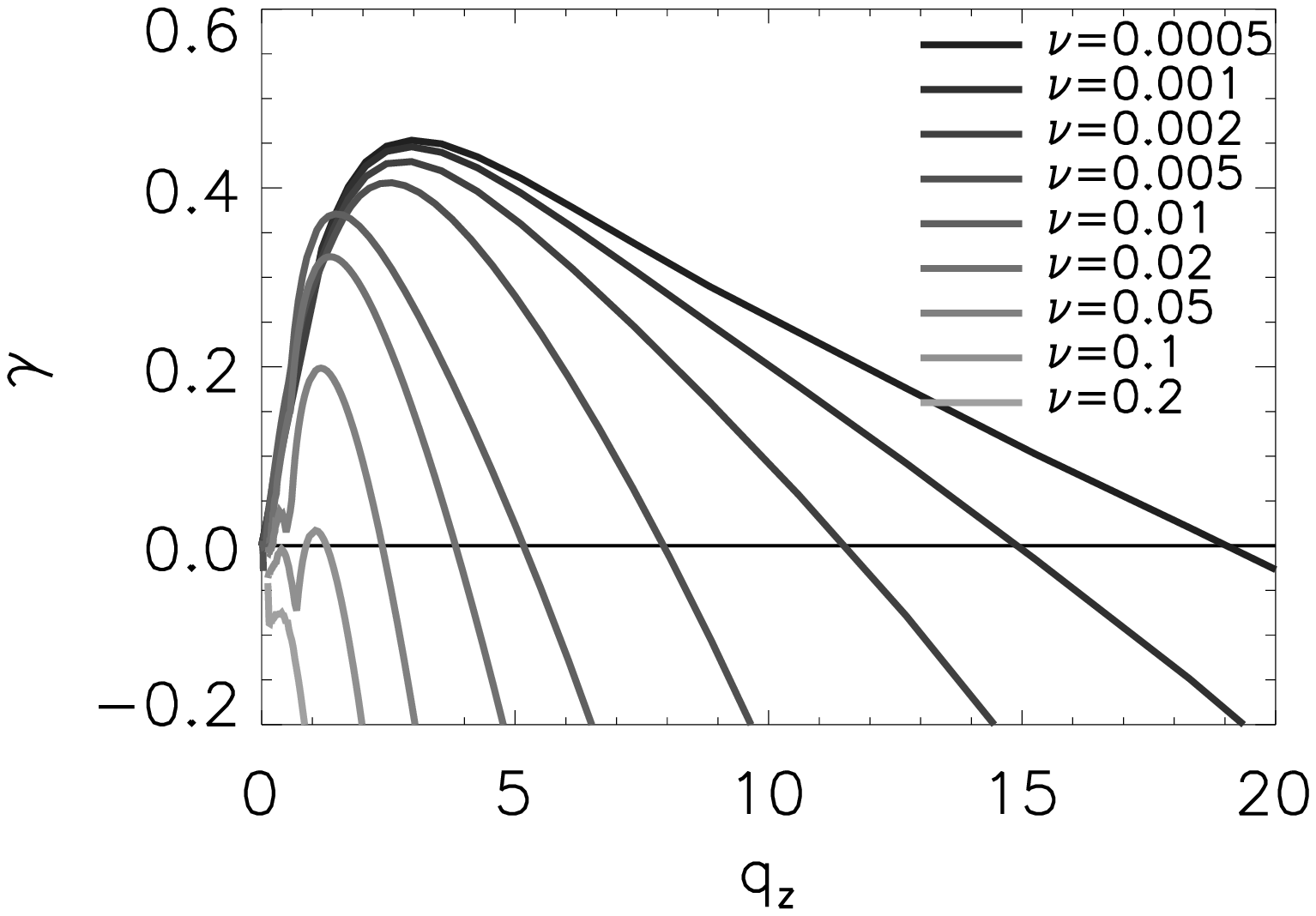}
  \includegraphics[width=0.45\textwidth]{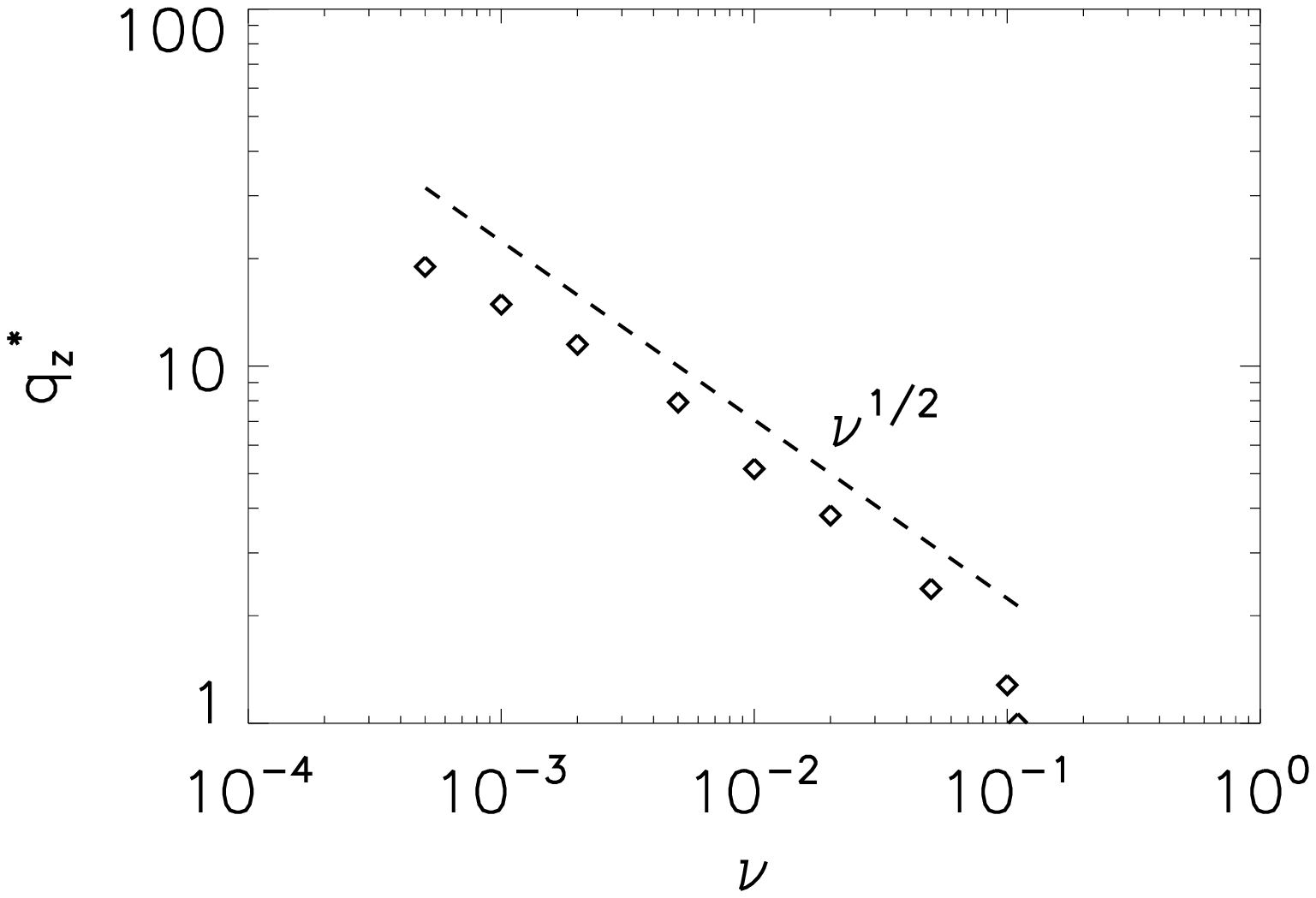}
  \caption{Left panel: The energy growth rate as a function of $q_z$. Right panel the critical wave number $q_z^*$ bellow which $q_z$ modes are unstable as a function of the viscosity. }
  \label{fig:KZ}
\end{figure}

Figure \ref{fig:KZ} displays the growth rate $\gamma$ as a function of $q_z$ for different values of the viscosity $\nu$. 
Above a critical value of $\nu>\nu_0\simeq 0.1$ modes with positive growth rates appear with $q_z\propto \ku=1$. As $\nu$ is decreased further
the range of unstable wave-numbers extends to larger and large values of $q_z$. For $\nu \gg \nu_0$ the unstable wavenumbers are found in the range $q_z < q_z^*$,
here $q_z^*$ is the largest unstable wavenumber. This critical wavenumber $q_z^*$ can be estimated from the graph in the left panel of figure \ref{fig:KZ} as the wavenumber for which the growth rate intersects the zero growth rate $\gamma=0$ line. It depends on $\nu$ and it is plotted  in the right panel of figure \ref{fig:KZ} in a logarithmic scale. 
It is shown to follow the scaling $q_z^* \propto \nu^{-1/2}$ for small $\nu$. This scaling is obtained from a balance of the viscous term $\nu {\bf q}^2 {\vs} $ in eq. (\ref{eq:floquet}) 
with the stretching term ${\bf v} \cdot \nabla {\bf U}$. A best fit results in 
\beq 
q_z^*\simeq 0.8 \left( \frac{U}{\nu\lu}\right)^{1/2} \label{qzcr}
\eeq 
where we have kept $U=1, $and $\lu=1$ in the expression to recover the dimensions.
For layers of finite height $H$ the presence of these unstable modes depends on $H$. 
This is because the periodicity of the domain in the $z$ direction imposes that the wavenumber of a 3D perturbation 
satisfies $q_z= n /H$ where $n\ge1$ is an integer. Therefore even for $\nu<\nu_0$ the flow will be stable to 3D perturbations if the layer thickness 
$H$ is thin enough so that $1/H$ is larger than the largest unstable wavenumber $q_z^*$. 


\subsection{Two and three dimensional perturbations }
\begin{figure}[!ht]
  \centering
  \includegraphics[width=0.55\textwidth]{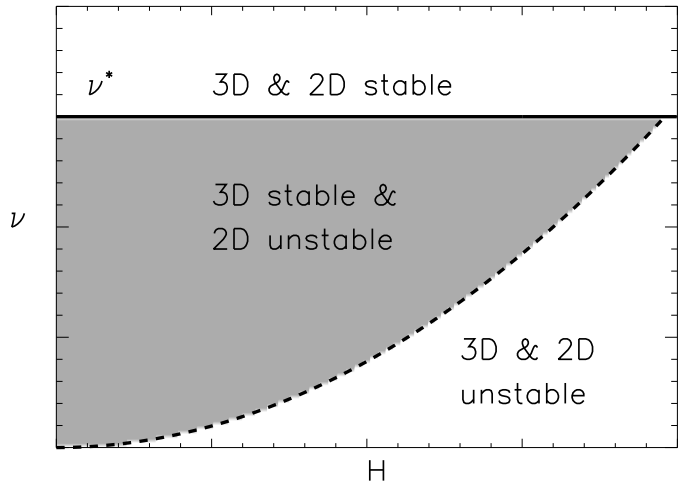}
  \caption{Stability diagram for a 2D flow}
  \label{fig:2Dinv}
\end{figure}
It is already possible to draw a first conclusion for 2D flows. 
From the analysis in section \ref{2D2Ds} it was shown that the large 2D modes are unstable if the viscosity is small enough 
\begin{equation}  \nu < \nu_*\simeq 0.7  U \lu \label{c1}  \end{equation}
so that the eddy viscosity for 2D perturbations satisfies $\nu_{eddy}+\nu<0$ (see eq. (\ref{nucr1})).
At the same time in section \ref{2D3Ds} it is shown that
3D modes are stable if the layer is thin enough so that $1/H$ is larger than maximum unstable wave number $q_z^*$ (see eq. (\ref{qzcr})). 
This implies stability of the 3D modes if 
\begin{equation} H \le \frac{1}{q_z^*} \simeq  \frac{\nu^{1/2} }{   0.8 (U \ku )^{1/2} }.  \label{c2} \end{equation}
Therefore there is a region is the parameter space $(H,\nu)$  determined by the conditions \ref{c1},\ref{c2} 
where 3D modes are stable while 2D modes are unstable to large scale perturbations by an eddy viscosity mechanism.
This is summarized in the diagram in figure \ref{fig:2Dinv} where condition \ref{c1} is given by the solid line while the condition \ref{c2} 
is marked by a dashed line. The region of 3D stable and 2D unstable modes is marked by gray.
In this region a transfer of energy to the large scales is expected and possibly an inverse cascade can develop. 
Of course whether an inverse cascade builds up or not depends on the nonlinear evolution of the unstable modes and can not be determined
by the stability of the laminar flow alone. Furthermore in the region that both 3D and 2D  instabilities are present we can not certify 
nor exclude a transfer of energy to the large scales. This is because we can not a priori conclude if the nonlinear evolution 
of the unstable 3D modes will suppress or not the inverse transfer by the negative eddy viscosity mechanism.

\section{Three dimensional flows }  

This next section considers the stability of 3D laminar flows given in eq. (\ref{eq:lam}) with in general $r\ne0$. For these flows the minimum layer thickness $H$ one can consider is given by the periodicity of $\bf U$ in the $z$ direction $H=\lu$.  Since the interest in this work is on thin layer for the remaining of this section we will fix $H$ to this minimum value. This limits the values of $q_z$ in the Floquet expansion to zero $q_z=0$ since all perturbations can only have vertical scales smaller than $H$. The investigation therefore will be restricted to large scale 2D perturbations with $\bf q$ of the form ${\bf q} = [q_x,0,0]$. It is noted that setting $q_z=0$ does not imply that ${\bf v}=e^{i\bf q x} \vs$ is independent of the coordinate $z$ because the field $ \vs$ depends in principle in all coordinates $x,y,z$ having the same periodicity as $\bf U$. Therefore 3D perturbations can still be unstable and as it is shown later they play an important role.

\begin{figure}[!ht]
  \centering
  \includegraphics[width=0.45\textwidth]{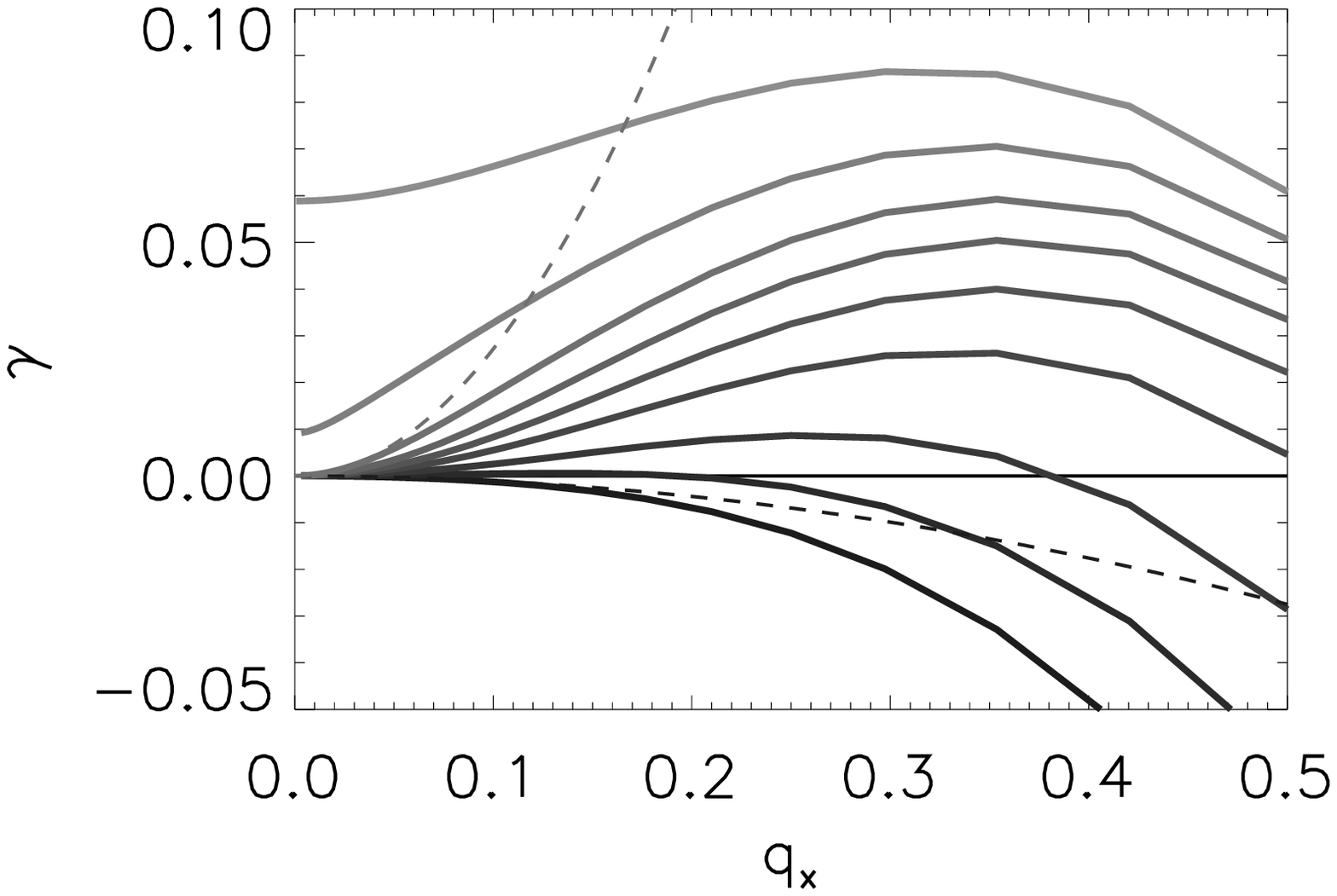}
  \includegraphics[width=0.45\textwidth]{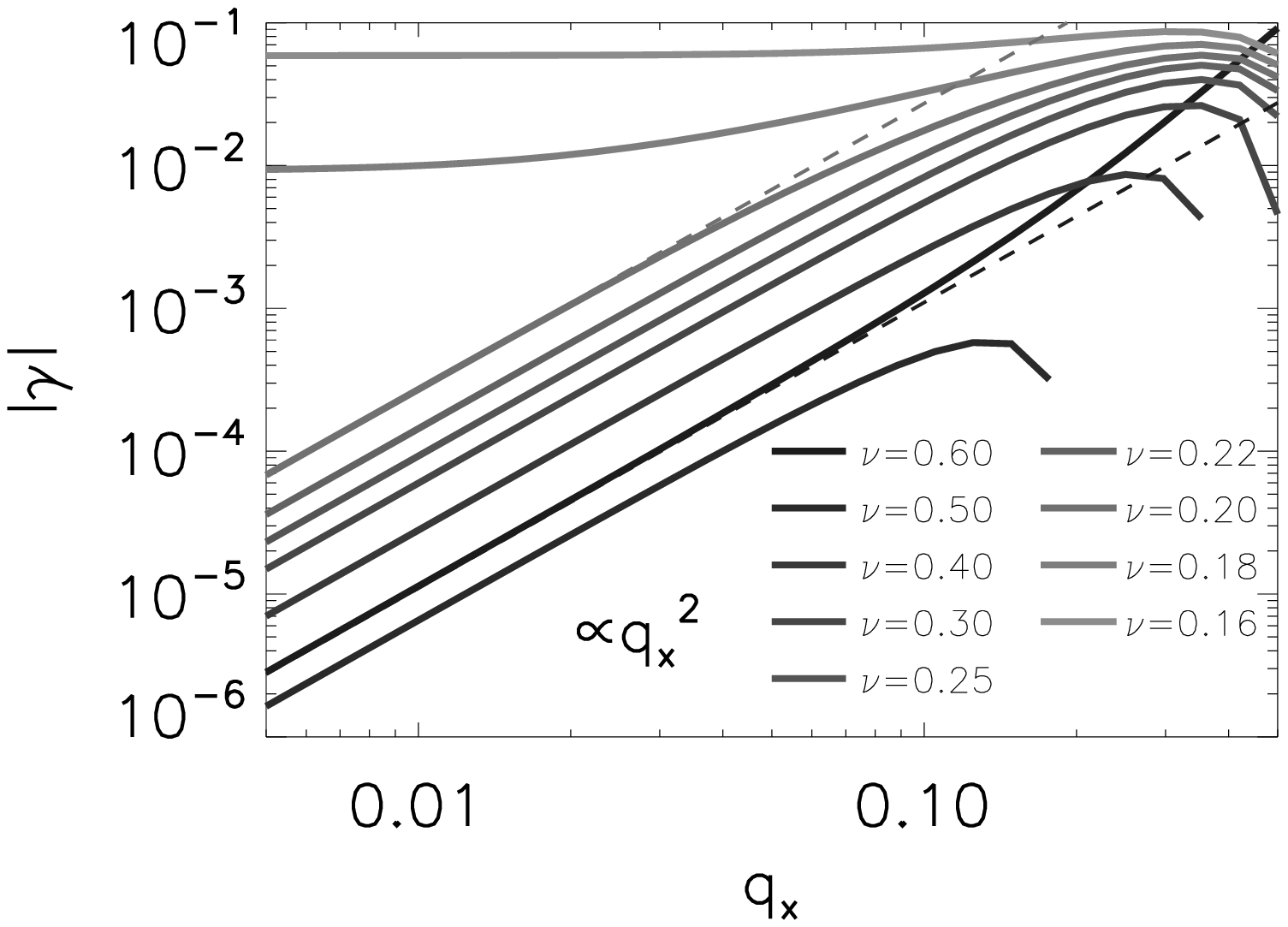}
  \caption{The growth rate $\gamma$ as a function of the wavenumber $q_x$ in linear scale (left panel) and logarithmic scale (right panel)
   for different values of $\nu$ and for $r=0$.}
  \label{fig:grth4} This work was supported by
\end{figure}

Figure \ref{fig:grth4} shows the measured growth rate as a function of the wavenumber $q_x$ for a flow with $r=0.4$ for different values of $\nu$. For small values of $\nu$ the growth rate shows a similar behavior as for the $r=0$ case shown in figure \ref{fig:2Dgrowth}. For $\nu$ bellow a critical value the growth rate is negative and with a quadratic dependence in $q_x$ as in eq. (\ref{q2}). Above this critical value the growth rate retains its quadratic behavior (eq.\ref{q2}) but with a positive value of the proportionality coefficient $a$. This behavior can then again be interpreted as an effect of negative eddy viscosity such that $\gamma = -(\nu+\nu_{eddy}) q_x^2$. However, if the growth rate is further increased the quadratic dependence on $q_x$ is lost and the growth rate attains a finite value a $q_x=0$.

\begin{figure}[!ht]
  \centering
  \includegraphics[width=0.45\textwidth]{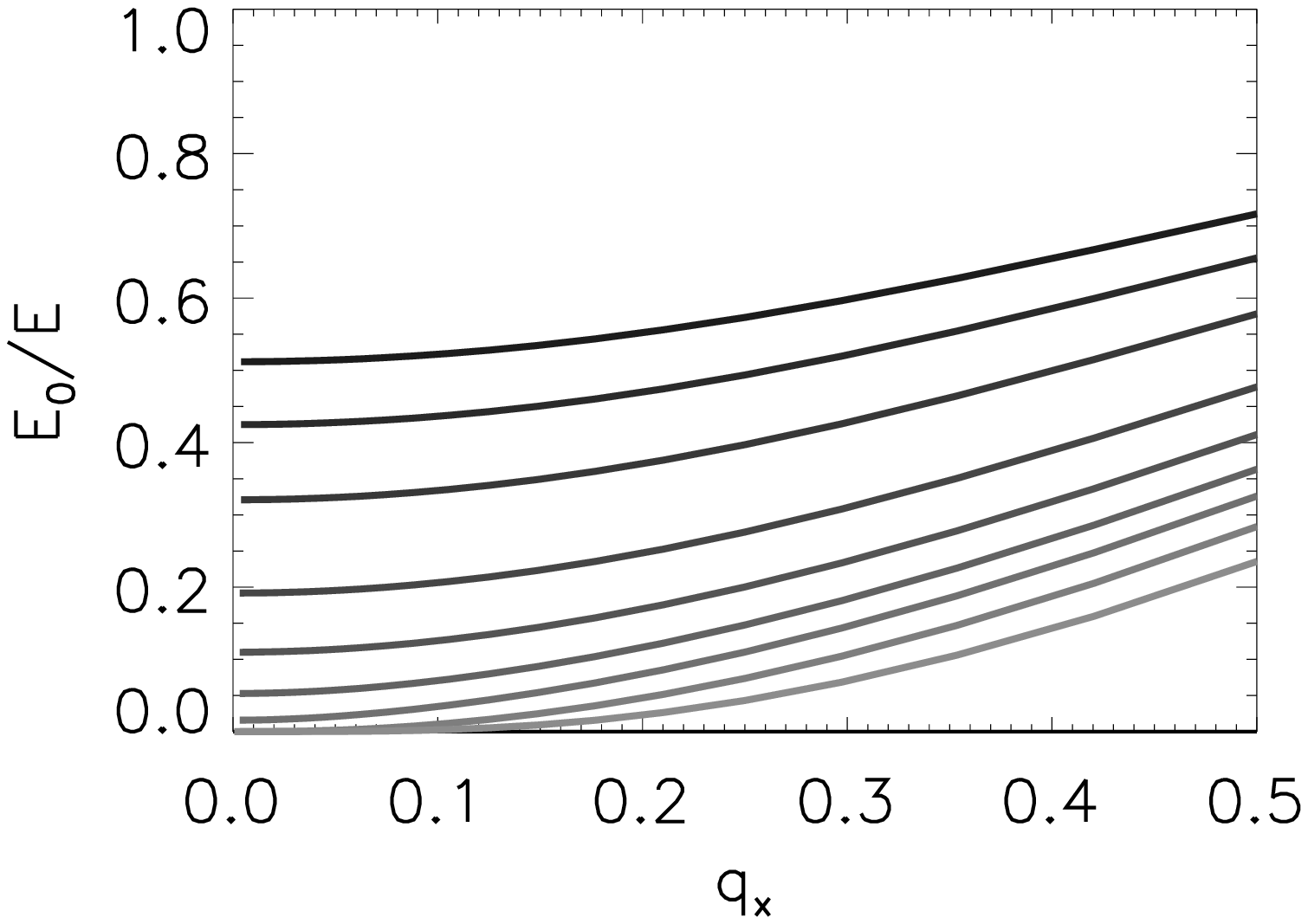}
  \includegraphics[width=0.45\textwidth]{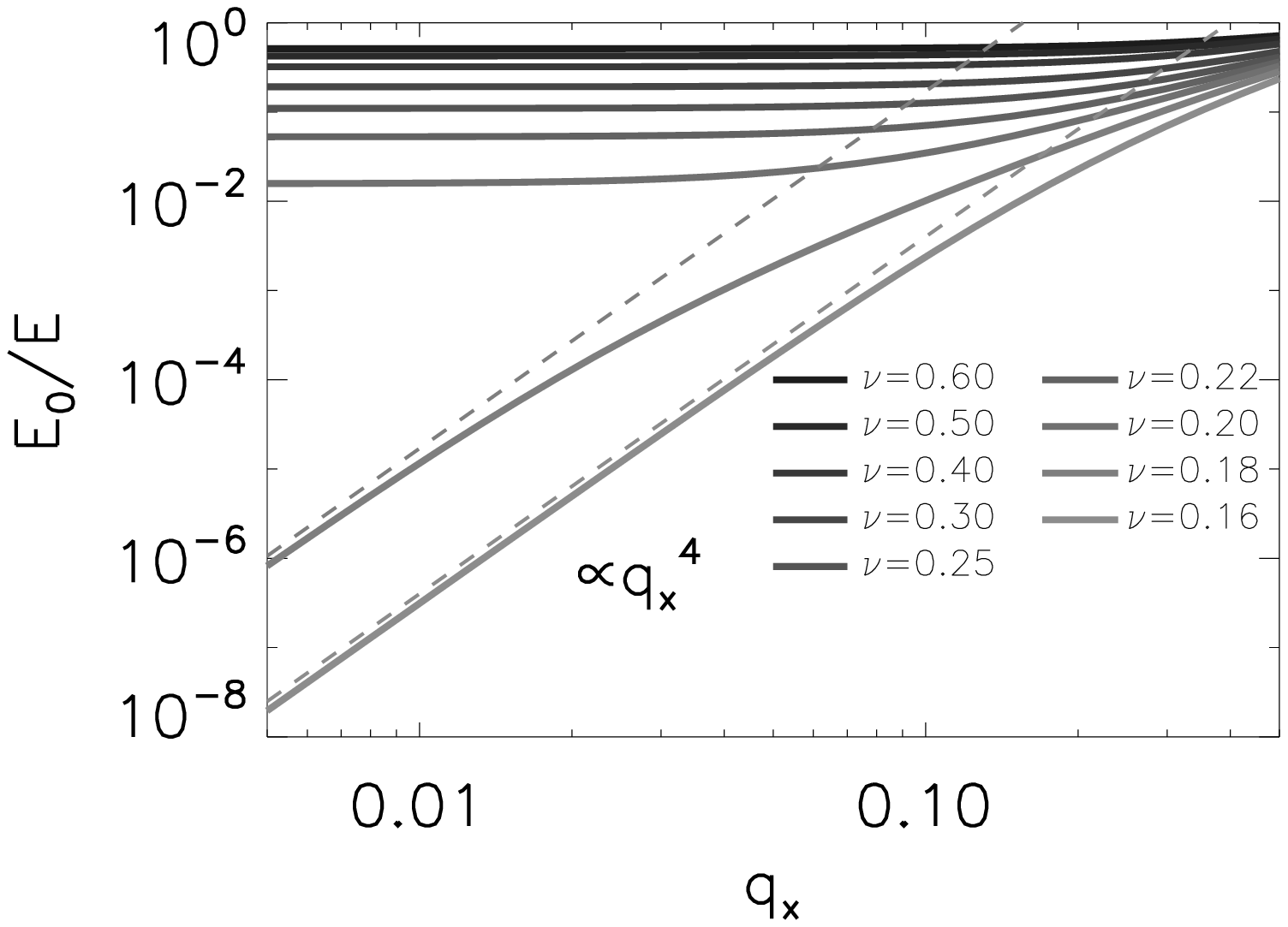}
  \caption{The ratio $E_0/E$ of the large energy $E_0=\frac{1}{2}|\langle \vs \rangle|^2$ to the total energy of the perturbation $E=\frac{1}{2}\langle |\vs|^2 \rangle$ as a function of $q$, for different values of $\nu$ in linear scale (left panel) and logarithmic scale (right panel). The color-shade of the lines is the same for both panels.    }
  \label{fig:Eratio4}
\end{figure}

A finite growth rate at $q=0$ can be a bit surprising since the growth of a non zero mean $\langle {\bf v} \rangle =\langle \vs \rangle \ne0$ field would violate momentum conservation. The contradiction is resolved by examining the relevant projection of the unstable mode to the large scales by looking at the ratio of  $E_0/E= |\langle \vs \rangle|^2/\langle |\vs|^2 \rangle $ that is shown in figure \ref{fig:Eratio4}.  This figure demonstrates that while for modes for which $\gamma \to 0$ as $q\to0$ the ratio $E_0/E$  remains finite, for modes with $\gamma \ne 0$ as $q\to0$ the projection to the large scales decreases with  $E_0/E \propto q^{-4}$. At $q=0$ therefore these last modes have a zero projection to the large scale modes, and correspond to purely small scales instabilities \cite{cameron2016large,cameron2016fate}.

Therefore the linear evolution of the large scale modes can be expressed in terms of an negative eddy viscosity up to this second critical value of $\nu=\nus$ bellow which 3D small scale instabilities begin. Large scale instabilities therefore are limited to values of the viscosity $\nu$ in the range $\nu_*>\nu>\nus$, where $\nu_*$ is the value above which large scales are stable, and for $\nu$ smaller than $\nus$ small scale instabilities are present. In principle for values of $\nu<\nus$ large scale modes ($0<q\ll1$) with positive growth rates $\gamma = -a (\nu+\nu_{eddy})q^2>0$ still exist but their growth rate is smaller than the growth rate of small scale instabilities of the $q=0$ modes. As a result for $\nu<\nus$ we can not predict if the flow would be able to transfer energy to the large scales with the linear evolution model given in \ref{LNS} and a nonlinear theory for the upscale transfer of energy is required.     

\begin{figure}[!ht]
  \centering
  \includegraphics[width=0.45\textwidth]{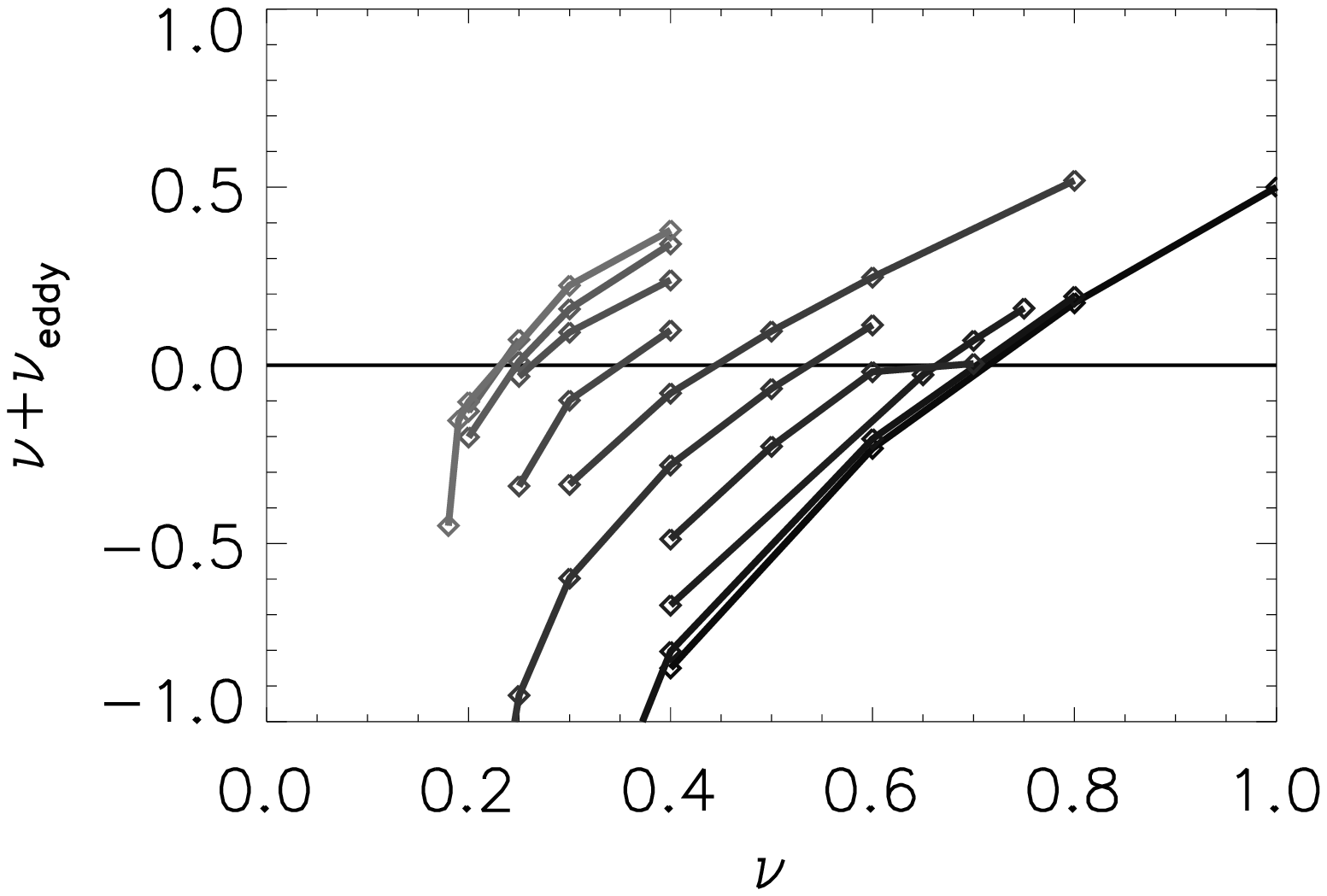}
  \includegraphics[width=0.45\textwidth]{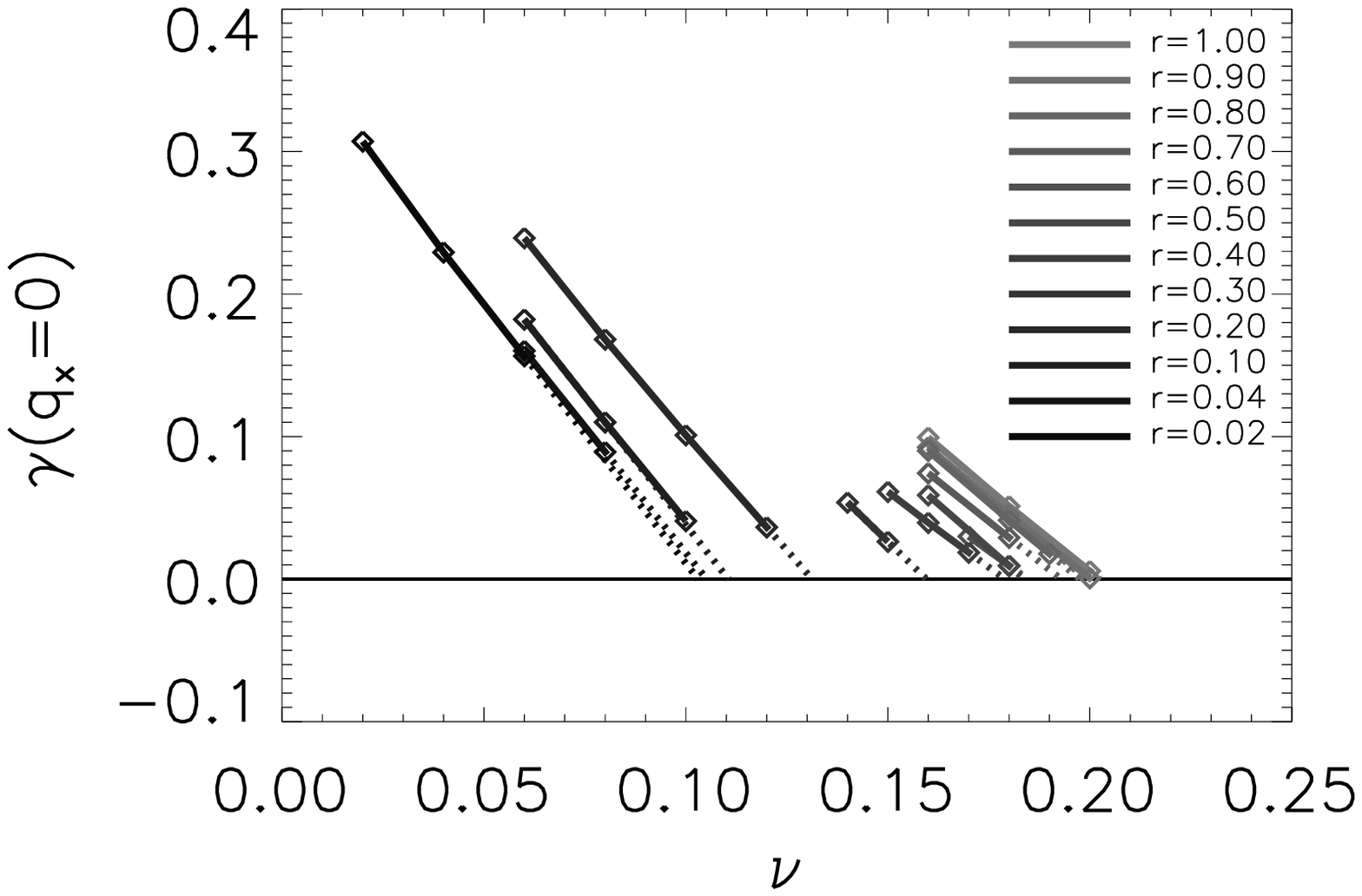}
  \caption{Left  panel: Effective viscosity as a function of $\nu$ for the different values of $r$ 
           Right panel: Small scale instability growth rate $\gamma(q=0)$ as a function of $\nu$ for different values of $r$.}
  \label{fig:nur}
\end{figure}

In the range $\nu>\nus$, the value of the eddy viscosity can be evaluated by fitting to a parabola as was done in section \ref{2D2Ds}.  
The values of the eddy viscosity for different values of $\nu$ and different values of $r$ were calculated and are plotted in 
the left panel of figure \ref{fig:nur}. 
This allows us to estimate the value of $\nu_*$ at which large scale 2D modes become unstable. 
Similarly we can calculate the value $\nus$ by looking at the growth rate of small scale instabilities $q=0$
and determining by extrapolation at which value of $\nu$ the growth rate $\gamma(q=0)$ of small scale modes becomes positive.
This is shown in the right panel of figure \ref{fig:nur} for different values of $r$. 

\begin{figure}[!ht]
  \centering
  \includegraphics[width=0.65\textwidth]{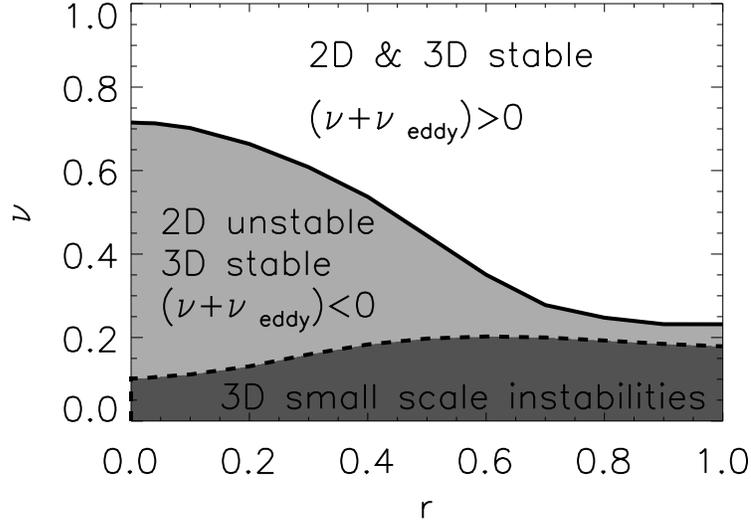}
  \caption{Stability diagram of the parameter space.  }
  \label{fig:rnu}
\end{figure}

The end result is shown in figure \ref{fig:rnu} where the stability diagram of the parameter space is shown in the parameter space $(\nu,r)$. 
The space is split in three regions. 
One (marked by white) for which the growth rate of all instabilities is zero,
one (marked by light gray) for which large scale modes are unstable by a negative eddy viscosity instability but 3D instabilities are zero,
and finally one region (marked by dark gray) for which small scale instabilities are present.
For 2D flows $r=0$ the region for which the effective viscosity is  negative, is larger. As $r$ is increased, the value of $\nus$ (dashed line) is increased and the value of $\nu_*$ (solid line) is decreased and as a result the range of values for which large scale instabilities can be observed is shrunk. Nonetheless even for $r=1$ for which the laminar flow does not contain any 2D part there is still a small range of values of $\nu$ for which only the large scale modes are unstable. This implies that transfer of energy to large scales can be observed in 3D flows as well.

\section{ Conclusions }

This work examined the linear stability of some basic linear flows against 2D large scale instabilities  3D small scale instabilities.
For thin layers and 2D laminar flows the onset Reynolds number for 3D instabilities increased as the layer thickness decreased. 
On the other hand large scale 2D flows become unstable due to an negative eddy viscosity type instability above a critical $Re$ independent
of the layer height $H$. This result implies that for sufficiently thin layers the 2D part of the flow can be unstable and transfer energy
to the large scales while 3D instabilities are suppressed. This result already indicates that depending on the layer height the flow can have different behaviors.
For small $H$ it can behave like a 2D flow cascading energy inversely while if $H$ is large the flow can be dominated by 3D instabilities 
cascading energy forward. The stability diagram  for the examined 2D flows is given in figure \ref{fig:2Dinv}.

For 3D flows it was shown that the negative eddy viscosity type of instabilities persist. There are therefore even for 3D flows unstable modes
that move energy to larger scales. These slow large scale instabilities however will be overwhelmed by 3D small scale instabilities if $Re$
is above their onset. For the examined flows is was shown that 2D instabilities have in general a smaller onset $Re$ than the 3D instabilities,
and thus a transfer to the large scales is still possible. However, as the flow becomes more 3D the difference between the two critical values decreases
leaving little room to observe a large scale instability. These results are summarized in figure \ref{fig:rnu}.

Finally we need to give a word of caution. The present results deal  only with the stability of the laminar flows.
The co-existence of of an inverse and forward cascade can not be neither confirmed or excluded from them.
This would require a study of the nonlinear state of the system that up to know can only be studied with the use of 
numerical simulations, and where the notion of eddy viscosity is not well established.   

\acknowledgments 

This work was granted access to the HPC resources of MesoPSL financed by the Region 
Ile de France and the project Equip@Meso (reference ANR-10-EQPX-29-01) of 
the programme Investissements d'Avenir supervised by the Agence Nationale pour 
la Recherche and the HPC resources of GENCI-TGCC-CURIE \& GENCI-CINES-JADE 
(Project No. A0010506421) where the present numerical simulations have been performed.
This work has also been supported by the Agence nationale de la recherche
(ANR DYSTURB project No. ANR-17-CE30-0004).

\bibliography{thin}

\end{document}